\documentclass[12pt,aasms]{myreport}
\raggedright
\newcommand{\emu}{$\rm \mu _{814}$}

\newcommand{\mo}{$\rm \mu _{0}$}

\newcommand{\mss}{mag arcsec$^{-2}$}

\newcommand{\plm}{$\pm$ }
\newcommand{\lb}{$\langle \:$}
\newcommand{\gb}{$\rangle \:$}

\newcommand{\gt}{$> \:$}
\newcommand{\lta}{$\leq $}
\newcommand{\gta}{$\geq $}

\newcommand{\Bmo}{\rm ${\mu _B}$(0) }

\newcommand{\alp}{$\alpha$\ }
\newcommand{\etal}{{et al}\ }
\newcommand{\degree}{$^{\circ}$ }

\newcommand{\app}{$\sim$}

\newcommand{\Msol}{$M_{\odot}$\ }
\newcommand{\Zsol}{$Z_{\odot}$\ }

\input{epsf}
\input{rotate}
\input{psfig}
\usepackage{rotating,caption2}

\begin{document}
\baselineskip=14pt
\def\oneskip{\baselineskip\baselineskip}
\pagestyle{myheadings}
\parskip=12pt
\def\oneskip{\baselineskip\baselineskip}
\floatsep 0.5in
\textfloatsep 0.5in

\setcounter{page}{1}
\thispagestyle{empty}
{\large \baselineskip=24pt
\begin{center}{\bf HST WFPC-2 Imaging of UGC 12695:  A Remarkably
Unevolved Galaxy at Low Redshift}
\end{center}
\vskip 0.2in
\centerline{Karen O'Neil, G.D. Bothun}
\centerline{Dept. of Physics, University of Oregon, Eugene OR, 97403}
\centerline{email:karen@moo.uoregon.edu}
\centerline{email:nuts@moo2.uoregon.edu}
\bigskip \bigskip
\centerline{C. D. Impey}
\centerline{Steward Observatory, University of Arizona, Tucson AZ, 85721}
\centerline{email:impey@as.arizona.edu}
\bigskip \bigskip
\centerline{and}
\bigskip \bigskip
\centerline{S. McGaugh}
\centerline{Physics Department, Rutgers University, Piscataway, NJ 08854 }
\centerline{email:mcgaugh@physics.rutgers.edu}}
\vskip 0.5in
Received \rule{2.0in}{0.01in}; accepted\rule{2.0in}{0.01in}\\
\clearpage
\markboth{\today}{\today}
\begin{abstract}
\setcounter{page}{2}
Utilizing the F814W and F300W filters, short exposure Hubble Space Telescope
Wide Field Planetary Camera-2 (WFPC2) images were taken of UGC 12695 
a nearby (z $\sim$ 0.021)
low surface brightness disk galaxy.  UGC 12695 has an unusual morphology,
consisting of a Y-shaped nucleus surrounded by a faint spiral arm with a
number of bright H II regions interspersed throughout the galaxy.
Surface photometry indicates the majority of
recent star formation in this galaxy occurred in these very localized regions,
most of which have a radius of \lta 2''.  This uneven stellar distribution
combined with the galaxy's overall extremely
blue color and low metallicity indicates UGC 12695 is an unevolved galaxy.

Some of the structural peculiarities
of this galaxy arise because a number of background galaxies, previously
thought to be morphological components of this galaxy, are showing through
both the outer nucleus and spiral arms of UGC 12695.  Surface photometry
of these galaxies shows them to be
fairly small ($\alpha\: <$ 1.8'') disk galaxies with total magnitudes ranging
from 19.6 through 24.2, and central surface brightnesses between
20.2 mag/arcsec$\rm ^2\:\leq\:\mu(0)\:\leq\:23.1\:mag/arcsec^2$.
When possible, the U, B, V, and I colors of these galaxies were measured
using ground-based images, which show the galaxies to be fairly red indicating
they are likely at redshifts z $\geq$ 0.5.
Inclusion of them in the photometry of UGC 12695 makes the galaxy
appear significantly redder.  When these galaxies are masked out,
the resultant U-I color of UGC 12695 is -0.2 $\pm$ 0.1, making it perhaps
the bluest galaxy every measured in this color system and confirming its
nature as a very unevolved galaxy at low redshift.
Spectroscopy of these
background galaxies, through the transparent disk of UGC 12695, may
help to clarify its chemical evolution and current heavy element content.
\vskip .2in
{\bf keywords} {galaxies: evolution; galaxies: star formation; galaxies: individual (UGC 12695);
galaxies: photometry; galaxies: spiral galaxies: stellar content}
\end {abstract}
\baselineskip = 24pt

\clearpage
\setcounter{page}{1}
\section{Introduction}
\vskip 0.25in
Studies of low surface brightness (LSB) galaxies have revealed a population
of relatively unevolved systems compared to the traditional Hubble
sequence of spirals (Bothun \etal 1997).  Although the overall
properties of LSB galaxies span a wide range in color, gas content, and
metallicity (see O'Neil \etal 1997; McGaugh \& de Blok 1997), there
is a significant subset that have extremely blue colors, high gas contents,
and low metallicities.  These are obvious candidates for being the most
unevolved systems in the nearby universe.  Among these kinds of systems,
UGC 12695 may be the best example of a very unevolved and quite
young LSB galaxy.

Previous studies of UGC 12695 have shown it to have a very blue color, 
low metallicity, high gas mass fraction, and unique morphological structure
(McGaugh \& de Blok 1997; McGaugh, Schombert, \& Bothun 1995; McGaugh
1994; McGaugh \& Bothun 1994; Klein, \etal 1992; Schneider, \etal 1990; 
Bottinelli, \etal 1990; Lewis 1987).
This galaxy is relatively isolated with its nearest neighbor (UGC 12687)
residing 11.2 arcminutes ($\sim$ 10 galaxy diameters) away.  Because
of its intriguing properties, UGC 12695 was a high priority target in our
Hubble Space Telescope (HST) Wide Field Planetary Camera-2 (WFPC2) survey 
of selected LSB galaxies to further study their structure and evolution.
In this paper, we report on the remarkable nature of this galaxy as 
revealed by these new observations. Section 2 describes the observations
and data reduction techniques. Section 3 describes
the global properties of UGC 12695, while section 4 discusses some of the individual
star forming regions of the galaxy.  In section 5 we discuss the background galaxies
discovered within the WFPC2 images, and finally in section 6 we consider the
implications of our discoveries.

\section{Observations and Data Reduction}

Eight HST WFPC2 images were taken of UGC 12695 on January 7, 1997,
using all four WFPC2 cameras.
The nucleus of UGC 12695 was positioned in the WF3 field. Initial image processing
occurred using the STSDAS pipeline, followed by additional reduction to
eliminate cosmic rays and then to combine the images. 

The WFPC2 consists of three Wide Field cameras and one Planetary camera.  The Wide Field
cameras have a focal ratio of f/12.9 and a field of
view of 80'' x 80'' with each pixel subtending 0.0996 arcsec.  The three cameras form an L-shape,
with the Planetary camera completing the square.  The Planetary camera
has a focal ratio of f/28.3, a scale of
0.0455 arcsec/pixel, and an overall field of view of 36'' x 36''.  Each camera has
an 800 x 800 pixel silicon CCD  with a thermo-electric cooler to suppress dark current.
The WFPC2 has two readouts formats -- single pixel resolution (FULL mode) and 2x2 pixel binning
(AREA mode).  All images taken through the F814W (814) filter were taken in FULL mode while
all images taken through the F300W (300) filter were taken in AREA mode.
 
Images of UGC 12695 were taken through both the 814 and the 300 filter.
The 814 filter is a broadband filter with $\lambda_0$ = 7924 \AA\ and $\Delta \lambda_{1/2}$
= 1497 \AA.   It is designed to be similar to the Johnson I-band filter.  The 300
filter has $\lambda_0$ = 2941 \AA\ and $\Delta \lambda_{1/2}$ = 757 \AA, and is designed to
be similar to the Johnson U-band filter.  All images had 500s exposure times.
As the sensitivity level through the 814 images was considerably higher than through the 300 
images (due to the CCD response), and UGC 12695 tended to be much brighter through the 814,
all galaxy identification and analysis was done with the 814 images.  When possible,
photometry for a galaxy was also done through the 300 filter.
 
Sky flat fields of the sunlit Earth were taken through each filter and routinely calibrated
against an internal flat field calibration system.  The internal system
consists of two lamps (optical and UV) illuminating a diffuser plate.
The internal flats are used to monitor and correct for changes in the
flat fields.  Dark frames are averages of ten calibration images taken over the space
of two weeks.  The intrinsic dark rate of the WFPC2 CCDs is \lta 0.01 e$^-$/pixel/sec.
The calibration dark fields were scaled to the
exposure time of each image.  A bias field was generated for each image using
extended register pixels which do not view the sky. 
 
The data reduction process was as follows: first, all known bad pixels were removed, using
the static mask reference file. 
The bias level, calculated as described above, was then removed
from each frame.  The bias image, generated to remove any position-dependent bias pattern,
was then subtracted from the image, as was the dark field image (described above).  Flat
field multiplication was then performed, using the fields described above.
All the preceding image calibration was performed at STScI using the standard WFPC2-specific
calibration algorithms (the pipeline).  
 
Four images of 500s each were taken through each filter.  After the images were reduced,
they were inspected for obvious flaws such as filter ghosts or reflections. 
Each frame was then shifted, registered and combined, using the STSDAS CRREJ
task to eliminate cosmic rays and other small scale flaws ($\sigma$ = 10,8,6). 
The resultant images were checked by eye to ensure any registration
errors were under 0.5 pixel.  The intensities were then reduced by a factor of 4
(using the IRAF IMARITH procedure) to give the resultant image
with mean intensity value from the four combined images.  Finally, the images were 
mosaicked, using the STSDAS WMOSAIC task to create a complete WFPC2 image
which is shown in Figure~\ref{fig:wfpc}.  The total flux was conserved in the
mosaic step.

Background galaxy identification was done within the confines
of the IRAF environment.  Each image
was enlarged by a factor of four and scanned
by eye for nonstellar objects. By examining both processed images available for each field
a minimum of four times a list was compiled of all possible nonstellar objects in each field
which had a minimum diameter of roughly 5 pixels (0.5'' for the Wide Field cameras, and 0.23'' for the
Planetary camera).  Automated galaxy search techniques were not employed for this image,
as one (FOCAS) had been used on other WFPC2 images and shown to be
less reliable than the by-eye search technique (O'Neil, \etal 1998).  All objects on the
list then had their appearance checked against their image in
one of the uncombined frames to insure no errors had occurred during the image
processing phase (e.g. image registration errors).  Remaining objects were considered
potential galaxies and left on the list.  It should be noted that because of the
low sensitivity in the 300 frames, all galaxy identification was done in the 814
frames.  Most of the galaxies were not visible in the 300
frames.

The zeropoints for each field were taken from the PHOTFLAM value given in the image headers.
The zeropoints from the STMAG system (the space telescope system based on a spectrum
with constant flux per unit wavelength set to approximate the Johnson system at V), is 
\[\rm ZP_{STMAG}\:=\:-2.5 log(PHOTFLAM)\:-\:21.1\]
For the 814 filter, the resultant zeropoint was 22.93, while for the 300 image the
zeropoints were 19.43 (for the PC image), and 19.46 (for the WF images).
Conversion of the zeropoints for the Vega system was done by comparing the
STMAG values of various objects within the field with the known values in both U and I
found from Michigan-Dartmouth-MIT 1.3m telescope images taken of UGC 12695.  (See McGaugh 1992
for details on the ground based images.)
The determined conversion factors were then (I $-$ 814) = $-$1.44 \plm 0.05,
(U $-$ 300) = 0.04 \plm 0.05.  
It should be noted that this conversion is from
the STMAG system to the Johnson-Cousins bands, not from the `Vega' system 
used by Holtzman \etal (1995), and others.  The difference in zeropoint 
conversions to Johnson-Cousins I-band
between these two magnitude systems is substantial for objects with the
colors of galaxies.   Appendix A provides more information and documentation
on this zeropoint difference.

The peak intensity for each galaxy was found and ellipses were fit around that point to obtain the
intensity in each annulus using the modified GASP software (Cawson 1983;
Bothun \etal 1986).  In the amorphous galaxies, like UGC 12695, 
the physical center, estimated by centroiding with respect to fitted isophotes, was chosen.  In the
cases of interacting (or overlayed) galaxies, the competing galaxy was masked, allowing
for a surface brightness profile to be obtained for each of the involved galaxies,
when possible.  The core of the point spread function has a radius of 0.1'' for the Planetary
camera, and 0.2'' for the Wide Field camera,
and the surface brightness profiles cannot be trusted below that.
The average sky-subtracted intensity within each (annular) ellipse was found and
calibrated with the photometric zeropoint.

Exponential surface brightness profiles were then derived using
\begin{equation}\rm \Sigma (r)\:=\: \Sigma_0\:e^{- r \over \alpha} \label{eq:sige}\end{equation}
where $\Sigma_0$ is the central surface brightness of the disk in linear units
(\Msol/pc$^2$), and \alp is the exponential scale length in arcsec.  This can also be written
(the form used for data analysis) as
\begin{equation}\rm \mu (r)\:=\:\mu (0)\:+\:({1.086 \over \alpha})r  \label{eq:mue}\end{equation}
where \mo\ is the central surface brightness in \mss.  

Galaxy inclination was found using the GASP software to determine the major
and minor axis at each isophote.  The inclination angle is 
\begin{equation}\rm {\it i}\:=\:cos^{-1}\left( {r_{minor}\over r_{major}}\right) . \label{eq:inclin}\end{equation}
Due to the asymmetric nature of most of these galaxies the use
of a flattening term ($q_o$) as part of the inclination derivation does
not seem appropriate.  We estimate that {\it i}
is accurate only to within \plm 5\degree when determined in this manner.
 
\section{Global Properties of UGC 12695}

With a diffuse, y-shaped nucleus and a faint spiral arm,
UGC 12695 has a unusual morphology, which is apparent
even from ground based images.  Additionally, this imaging
has shown what appears to be a number of bright star forming knots 
spread throughout the galaxy (Figure~\ref{fig:gnd}).  Morphological
classification on these images would suggest that UGC 12695 is not
a well-defined spiral and would therefore most likely be classified
as a Type I Irregular galaxy, with sporadic locations of star formation.
We emphasize, however, that UGC 12695 is considerably more massive
than the prototypical examples defined by the LMC and NGC 4449.
Furthermore,
the HST data has shown that appearances are deceiving as most of these
``star forming knots'' are, in fact, background galaxies clearly shining
through the main body of UGC 12695.  This discovery changes
our view of this galaxy considerably.

Table~\ref{tab:glob} lists the global properties
of UGC 12695 found primarily from 
previous studies.  In the first row, columns 1 and 2 give the coordinates (RA and Dec) in 
the J2000 coordinate system found using the WFPC2 image and the STSDAS METRIC task.
Column 3 lists the heliocentric radial velocity obtained from the 21-cm
observations of Bothun \etal (1985). 
Column 4 gives our assumed distances based on a Hubble Constant of 75 Mpc/km/s
and a Virgocentric infall velocity of 300 km s$^{-1}$.
Column 5 gives the total HI mass of the galaxy, with $\rm M_{HI}\:=\:log (M_{HI} / M_{\odot})$
and column 6 gives the total dynamical mass of UGC 12695
($\rm M_{dyn}\:=\:log(M_{dyn}/M_{\odot}$)).
In column 7 is listed the gas mass to luminosity ratio, while column 8 provides the 
gas mass fraction for the
galaxy, with $\rm f_g\:=\:{{M_g}\over{M_g\:+\:M_*}}$.  (Values listed in
columns 6--8 come from McGaugh 1992.)
Column 9 lists the exponential scale length in kpc 
(McGaugh \& Bothun 1994).  In column 10 is the galaxy's absolute magnitude (McGaugh 1992).  Column
11 lists the galaxy's central surface brightness, determined from equation~\ref{eq:mue}
(McGaugh \& Bothun 1994).  Finally, columns 12 - 15 give the galaxy's apparent magnitude and 
total (luminosity weighted) colors, also from McGaugh \& Bothun (1994).

Examination of Table~\ref{tab:glob} shows that UGC 12695 is quite blue
and rich in gas, suggesting that it is now undergoing one of its first
episodes of star formation.  The chance to study this phenomena at
significantly higher resolution motivated the HST observations.
The first surprising discovery from the WFPC2 image of UGC 12695 was that many
of the objects believed from ground based observations to be stellar knots contained
within UGC 12695 are actually background galaxies which are showing {\it through}
both the galaxy's disk and outer nucleus.  The significance of this discovery 
is two-fold -- 
first the number of putative star forming regions has been greatly reduced
and second UGC 12695 is a remarkably transparent galaxy, even in its nuclear
regions.
As all prior photometry of
UGC 12695 was done assuming the background galaxies to be part of UGC 12695, we
re-determined the galaxy's surface photometry with the galaxies masked (and the 
masked pixels replaced with
the average values of the surrounding pixels).  

The results of re-determining UGC 12695's photometry with the background galaxies masked are shown in
the first two rows of Table~\ref{tab:col}.  In the first row (a) we show the colors
derived for UGC 12695 from ground based images, with
the background galaxies included in the photometry, while row b shows the colors of
UGC 12695 once these galaxies are masked.  (See Table~\ref{tab:backg} to see which galaxies
were masked during this process and which are too far from the nucleus of UGC 12695 to have been 
included in any of the photometric calculations.)  The significance of masking
the background galaxies is immediately apparent -- masking the background
galaxies  (which are primarily red, see below) has made this already very
blue galaxy significantly bluer.  In fact, the resultant U-I color of 
\app -0.2, may make UGC 12695 the bluest known galaxy.  It
is remarkable that this blue color is achieved without many obvious large scale 
regions of star formation being present.  It is thus the smooth, diffuse light
of this galaxy, coupled with small scale regions of
star formation, which is very blue.  This assertion is documented below.

Taking advantage of the high resolution of the WFPC2 images, we further explored
UGC 12695's global properties by masking the six H-\alp
star forming regions listed in Table~\ref{tab:hcol} and determined from 
ground based images taken at the Michigan-Dartmouth-MIT 1.3m McGraw-Hill telescope. 
Row c of Table~\ref{tab:col}
shows the colors of UGC 12695 with both the known background galaxies and the H-\alp
regions removed.  Significantly, no measurable difference between rows b and c can be noted.
Additionally, Figure~\ref{fig:sbprof} shows the (814 band) surface brightness profile
of UGC 12695 with all background galaxies and H-\alp regions included in the 
analysis (solid line), UGC 12695 with all the definite background galaxies removed 
(dashed line), and UGC 12695 with both the background galaxies and all the H-\alp regions
removed (dash-dotted line).  No differences are apparent in the plots until after r=20'',
and no difference in the overall profile of the galaxy can be seen at any radius.

The discovery of background galaxies shining through UGC 12695 dramatically impacts the
galaxy's nuclear (inner 10'') color.  It should be noted, though, that as these
galaxies appear primarily through the 814 (I band) filter, and mostly `dropped out'
when viewed through either the (ground based) B band or the 300 images, the 
removal of the galaxies does not
affect the overall magnitude, surface brightness, gas fraction, etc. of UGC 12695
listed in Table~\ref{tab:glob} and found using the blue luminosity.
Hence, the primary effect of the mistaken inclusion of these background
galaxies in prior studies is in the total I-band luminosity and the
derived U-I or V-I colors.  Removal of these galaxies significantly
lowers these values.

\section{Star Formation in UGC 12695}

The high resolution of the HST WFPC2 cameras allows for an unprecedented view of the
stellar structure of UGC 12695.  Examining the 814 image of UGC 12695
(Figure~\ref{fig:wfpc}(a)) shows a number of distinct star forming regions surrounding
a fairly diffuse, irregular nucleus.  Additionally, a spiral arm can be seen encircling the nucleus, which
consist of disjoint stellar knots.
Figure~\ref{fig:nuc} shows a greyscale image of the inner 10'' of UGC 12695.
Highly non-circular in appearance, the nucleus of UGC 12695
consists of distinct regions of star formation clustered primarily in the northern (lower)
portion.  Other than these localized regions of star formation, even the nuclear regions
of UGC 12695 are transparent, to the point were a spiral background galaxy can be identified
clearly (bottom edge of Figure~\ref{fig:nuc}).  We do note, however, that
this nuclear region is significantly redder than the galaxy as a whole
(see Table 2) and thus likely contains the bulk of any old stellar
population that might be present.  The LSB nature of this galaxy, however,
strongly indicates that this older population is distributed in 
a very diffuse manner and is certainly not concentrated into a central
bulge or bar.

A number of regions of H-\alp emission are also present within UGC 12695, many of which are visible
in the ground based images (Figure~\ref{fig:gnd}e).  Photometry of these areas is
given in Table~\ref{tab:hcol}, with the detailed areas studied shown in
Figure~\ref{fig:upict}.  The majority of these areas are fairly
small, with r$_{27}$ \lta 2.6'' (0.8 kpc).  (The only exception is area E, which has a radius
of 6.4'' (2 kpc).)  The colors of these regions are extremely blue, with
0.1 \lta\ (U-I) \lta\ 0.5, indicating a significant level of current star formation.

McGaugh (1992) performed a metallicity study on a number of UGC 12695's H-\alp regions.  
The slit positions used are shown as white lines in Figure~\ref{fig:upict}, while the
results from that work are given in Table~\ref{tab:metal}.
Slit one (s1) runs east-west through the northern H-\alp
areas, while slit two (s2) runs roughly north-south through the southern areas.
The areas labeled s2a3 (unfortunately) lies on an H-\alp area which is overshadowed
in the 814 image by a bright background galaxy.  The metallicities of these areas
are low, around 0.4\Zsol.  Combined with the extremely blue colors
of the regions of H-\alp regions, this indicates that these regions are
likely young starburst areas which have not yet enriched much of the
gas.

A further look at two of these regions of H-\alp emission (areas B \& C) is given in Figure~\ref{fig:halp}. 
We have
determined the 814-300  colors of localized regions within these H-\alp areas, with
the results listed in Table~\ref{tab:halp}.  Regions 1 - 8 within the H-\alp area
were chosen to encompass various regions of localized starburst.  The colors of these
regions are exceptionally blue, and thus are likely comprised primarily of young
stellar types.  Significantly, though, region 8 which, although clearly associated with
the H-\alp areas, does not encompass a regions of apparent significant star forming
activity, also has an 814-300 color which is at least 1 magnitude bluer than UGC 12695's
nuclear region.  It is thus clear that the majority of the current star formation in
UGC 12695 is happening within very localized (small spatial scale), 
non-centralized, star forming regions which
are dispersed throughout the galaxy.

\section{The Background Galaxies}

As these are short exposure images of the nuclear 
region of a nearby galaxy, the existence of any background galaxies 
in our image was a large surprise.  The fact that we discovered
21 potential background galaxies is proof both of the fairly
transparent nature of UGC 12695 as well as the excellent ability of
HST WFPC observations to detect distant galaxies no matter where its
pointed.

Details of the galaxy identification and analysis techniques are given in section 2 of this
paper, while the results, as applied to these background galaxies,
are listed in Table~\ref{tab:backg} and described below.  The 814 band detection limit was 24.5 mag.

\begin{list} {\setlength{\rightmargin 0.5in}{\leftmargin 0.5in}}
\item { Column 1:} The name given to each of these galaxies (none have been previously
identified).   The numbering system is based on all objects identified in the image, including
foreground stars, image flaws, and regions of H-\alp emission.
\item { Columns 2 and 3:} RA and Dec in J2000 coordinates, determined using
the STSDAS METRIC task. 
\item { Column 4:}  The 814 - 300 color of the galaxies.  If the galaxy could not be identified
in the 300 image, a maximum apparent brightness (magnitude) was determined through assuming the 
galaxy would have been identified if its radius exceeded 5 pixels and it had an average brightness
greater than 3$\sigma$ above the sky. 
(Sky brightness was determined for each of the four WFPC2 frames separately.). 
\item { Column 5:} The U-I color of each galaxy, as described in section 2. 
\item { Columns 6 and 7:} The total magnitude of the galaxy through both
the 814 and 300 filters.   Magnitudes are corrected for galactic extinction (treating
the 814 filter as a Johnson I band filter and the 300 filter as a Johnson
U band filter) but not for inclination or redshift (k-correction). 
\item { Column 8:} The total integrated magnitude of the galaxy through the 814 band
using \begin{equation}\rm mag(\alpha )\:=\: \mu (0)\:-\: 2.5 log(2\pi \alpha^2) \label{eq:malph}\end{equation}
where \alp is the exponential scale length in arcsec.  If an exponential profile was not
fit to a particular galaxy's surface brightness profile this column is left blank. 
It should be noted that on occasion (i.e. U2-18) the magnitude in this column is considerably brighter
than that given in column 7.  In these cases the maximum aperture size used in calculating the
total magnitude (column 7) was smaller than the actual galaxy size, typically due to interference from
a neighboring galaxy or from UGC 12695.  The integrated magnitude given in this column is thus the more accurate.
\item { Column 9:} The exponential scale length of the galaxy, found as described in 
section 2, equation~\ref{eq:mue}. 
\item { Column 10:} The central surface brightness of the galaxy in \mss,
as defined in section 2, equation~\ref{eq:mue}. 
When a surface brightness profile was found for a galaxy,
yet no line was fit to it, the central surface brightness is estimated. 
\item { Column 11:}  The inclination corrected central surface brightness in \mss.
\begin{equation}\rm \mu_e (0)\:=\:\mu (0)\: -\: 2.5log(cos({\it i})), \end{equation}
where the inclination used is listed in column 13.
Note that this is a geometric path length correction which assumes no dust, and therefore
may not be accurate for these galaxies.  
\item { Column 12:}  The major axis radius in arcsec as measured at the \emu = 25.0 \mss\
 isophote.  If the surface brightness profile errors exceeded 0.25 \mss\  before \emu
= 25.0 \mss, then the largest accurate radius is given.
\item { Column 13:} The inclination angle (in degrees) as found by the GASP
software (equation~\ref{eq:inclin}).  The angle is correct to within \plm 5\degree.  (The
error is primarily due to the galaxies' lack of azimuthal symmetry).
\end{list}

Eight of the galaxies were large and bright enough to be identified within the MDM
images.  These galaxies are listed in Table~\ref{tab:gndgal}, along with the galaxies'
colors, as determined from the ground-based images.  
As these are distant galaxies, none could be reliably identified through the B band image,
and a few could not even be identified in the V band image.  In these cases the maximum apparent 
magnitude was determined for the `drop-out' band, as described above, using a minimum
detection radius of 3 pixels.

The majority of the galaxies discovered have sizes \lta 1.5'', making a morphological 
description difficult at best.  A few of the galaxies, though, are fairly large and thus 
could be examined more closely.   Figure~\ref{fig:backg} shows the four largest background 
galaxies discovered in our image.  All four appear to be well-formed spiral galaxies, with
little of the irregular morphology evident in deeper HST WFPC2 surveys such
as the Hubble Deep Field (Williams \etal 1996).    Followup spectroscopy of
these galaxies is clearly desirable.

Comparing the values of these background galaxies with the colors, sizes, and magnitudes of
both the medium and deep HST surveys leads to the conclusion that the background galaxies in
our survey lie at a redshift of 0.5 \lta z \lta 1.5.  The radii of the galaxies in the
medium-deep survey (MDS) lie between 1.2'' -- 2.2'', while the V $-$ I color of the
MDS galaxies is 0.1 \lta V $-$ I \lta 3.0, with \lb V $-$ I \gb = 1.2, and the survey ranging from
20.0 \lta m$_{814}$ \lta 22.0.  The background galaxies in our survey thus lie at the
red edge of this color spectrum and are typically smaller (in apparent size) than those in the
MDS.  As the MDS galaxies were determined to lie between 0.5 \lta z \lta 1.0, we conclude
the background galaxies in our survey lie at least that distant, or at 0.5 \lta z \lta 1.5. 
In order to more fully understand the morphology and distribution of these galaxies, and 
to thereby gain a firmer understanding of galaxy morphological evolution, 
a more detailed look at these galaxies through multiple filters and
at longer exposure times, is definitely warranted.

\section{Discussion and Conclusion}

UGC 12695 is a remarkable galaxy.  It has an exceptionally high gas mass 
fraction, very low metallicity, a strikingly diffuse morphology with 
perhaps the bluest colors known for a galaxy, and an exceedingly transparent 
nature.  Combined, these attributes indicate UGC 12695 is a highly unevolved galaxy. 
Since UGC 12695 is at rather low
redshift (z$\sim$0.021) then its properties indicate that some potentials may
well have late collapse and formation timescales (such as may also be
the case in NGC 1705, another apparently young galaxy - see Meurer \etal
1992).  The discovery of apparently young galaxies at low redshift has
considerable importance to understanding galaxy evolution.  As its
unlikely there is anything unique about UGC 12695, one would reasonably
expect objects like it to exist at any redshift.  However, the LSB
nature of UGC 12695 makes it difficult to detect.  Indeed  it seems
ironic that an apparently unevolved galaxy would itself difficult to
detect because of its low surface brightness.  This obviously has
important implications for detecting young galaxies at any redshift.

Understanding the evolution of UGC 12695 might shed considerable light into 
galaxy formation scenarios as a whole.  Unlike most galaxies, UGC 12695 also 
has the advantage of being relatively isolated, with its only nearest neighbor 
located at a projected distance of \app 200 kpc. 
UGC 12695 is likely to have been influenced minimally (if at all) by other galaxies. 
Clues to the formation of UGC 12695 may be found in its structural 
peculiarities.  The galaxy's low overall density is manifest in its dynamical
mass-to-total size, as well as its transparent nature. 
The inferred low mass density of UGC 12695 (and LSB disks in general -
see de Blok and McGaugh 1997) may mean that stellar orbits are
not well-defined in this potential thus resulting in the observed
non-circular nature of even the inner nucleus and the fact that the 
majority of the star formation is taking place in non-centralized, 
yet strongly localized regions spread throughout the diffuse galaxy body.   
Whatever is the case, the overall properties indicate that UGC 12695
is still highly unevolved.  The clear fact that this object
has escaped our attention, due to its diffuse nature, causes concern over
how many other examples of this phenomenon in the nearby universe have
been missed to date.

This research has made use of the NASA/IPAC Extragalactic Database (NED) which is operated
by the Jet Propulsion Laboratory, California Institute of Technology, under contract with
the National Aeronautics and Space Administration. 

\clearpage
\centerline{Appendix A}
\centerline{Photometric Calibration of HST WFPC2 Images}

The calibration we have used in converting our 814 data to the Johnson-Cousins
system is substantially different from the values given in Holtzmann
\etal (1995).  The primary reason for this is because our instrumental
zeropoint is based on the ST magnitude system while the Holtzmann \etal
values use Vega to define the zeropoint.  As detailed by Whitmore  in the
WFPC2 photometry manual (STSCI internal publication) the difference
in conversion factors is substantial.  This is because the 814 filter
is a close approximation to Johnson-Cousins I, but the definition of
ST mags, for a star of color corresponding to K0III (a reasonable
approximation to the SED of a galaxy) is different by 1.2 magnitudes
than if Vega were used as the zeropoint.   Since this difference may not be
widely known in the community, unless the WFPC2 photometry documentation
has been read carefully (specifically Chapter 41), 
we give a brief synopsis here.

Both the 300 and 814 filters are
wide bandpass filters, designed to maximize the throughput, and were not
designed to specifically match a standard filter bandpass.  
The STMAG system exists with
the true filter design in mind and is based on a spectrum with constant flux per unit
wavelength.  Conversion from the DN to the STMAG system is thus straightforward:
\[\rm STMAG\:=\:-2.5log\left\{ DN/(exposure\: time)\right\} \:-\:2.5log(PHOTFLAM)\:-\:21.1\]
where PHOTFLAM is a value obtained by STScI using standard star calibrations with the
four WFPC2 chips, and takes into account distortions, plate scale, etc.
Because the STMAG system is not a standard magnitude system, though, 
conversion from the DN is often instead done to a more conventional system based on
Vega's spectrum (i.e. Holtzmann \etal 1995; Driver, Windhorst, \& Griffiths 1995). 
Conversion from the STMAG system to the Vega system
is typically done through 814$_{STMAG}$ $-$ 814$_{Vega}$ $\approx$ -1.226,
300$_{STMAG}$ $-$ 300$_{Vega}$ $\approx$ -0.05.  Whichever of these systems is chosen, though,
additional corrections need to be made to convert the data to Johnson-Cousins colors.

Before discussing the additional corrections necessary for conversion to the
Johnson-Cousins system, an additional factor needs to be considered.  WFPC2 images 
can be taken with two different gain settings, 7 e$^-$/ADU and 14 e$^-$/ADU.
The calculations done by Holtzman \etal (1995) are for the gain=14 setup,
while the data contained in this paper, as well as for the majority of 
the recent WFPC2 data was taken with the gain=7 setting.  This results in
a difference of roughly -2.5log(2) = -0.75, but of course has a further
dependence on the spectral energy distribution of the source.

Our ground-based I-band data, which is used to convert the 814 data to
the Johnson-Cousins system, was obtained through the standard KPNO
filter set.   This I-band filter has  $\lambda_{peak}$ = 8290 $\pm$ 40
\AA\ and  $\Delta \lambda$(FWHM) = 1950 \AA.   The traditional Johnson
I-band filter has $\lambda_{peak}$ 9000 \AA\ and $\Delta \lambda$(FWHM) = 
2400 \AA.  Because CCDs have considerably more red sensitivity than the
old S-20 phototube, the peak wavelength of the I-band filter used for
CCD observations is significantly bluer.  The 814 filter used in the
HST observations has 
$\lambda_{peak}$ =
8386 \AA, \lb$\lambda$\gb = 8269 \AA, $\Delta \lambda$(FWHM) =  1758 \AA,
values that are similar to the I-band filter used from the ground.  Had
both the ground-based and HST data been calibrated against Vega, we
would have recovered the Holtzmann \etal (1995) conversion.  The
much larger conversion that we found (e.g. 1.4 mags) is due almost entirely
to the use of the STMAG system for our WFPC2 data with a gain setting of 
7 instead of 14 (again this is all discussed in Chapter 41 of the WFPC2
photometry manual). 

\clearpage
\centerline{References}

Bothun, G., Impey, C., \& McGaugh, S.  1997, PASP 109, 745\\
Bothun, G.D. Mould, Jeremy R., Caldwell, Nelson, \& MacGillivray, Harvey T.  1986, AJ, 92, 1007\\ 
Bothun, G.D., Beers, T., Mould, Jeremy R., and Huchra, J. 1985, AJ 90, 2487\\
Bottinelli, L., Gouguenheim, L., Fouque, P., \& Paturel G.  1990, A\&AS, 82, 391\\
Cawson, M.  1983, Ph.D. thesis, University of Cambridge\\
Driver, Simon P., Windhorst, Roger A., \& Griffiths, Richard E.  1995 ApJ 453, 48\\
de Blok, W.J.G., \& McGaugh, S.  1997, MNRAS, 290, 533\\
Holtzman, J., Burrows, C., Casertano, S., Hester, J., Trauger, J., Watson, A., \& Worthey, G.
1995 PASP 107, 1065\\
Klein, U., Giovanardi, C., Altschuler. D.R., \& Wunderlich, E. 1992, A\&A, 255, 49\\
Lewis B.M. 1987, ApJS, 63, 515\\
McGaugh, S.S., 1994, ApJ, 426, 135\\
McGaugh, S.S., 1992, Ph.D. thesis, University of Michigan, Ann Arbor\\
McGaugh, S.S., \& de Blok, W.J.E. 1997, ApJ, 481, 689\\
McGaugh, S.S., Schombert, James E., \& Bothun, G.D. 1995, AJ, 109, 2019\\
McGaugh, S.S., \& Bothun, G.D. 1994, AJ, 107, 530\\
Meurer, Gerhardt R., Freeman, K. C., Dopita, Michael A., \& Cacciari, Carla 1992, AJ, 103, 60\\
O'Neil, K., Bothun, G.D., \& Impey, C.D. 1998, in preparation\\
O'Neil, K., Bothun, G.D., Schombert, J., Cornell, M. \& Impey, C.  1997 AJ, 114, 2448\\
Schneider, S.E., Thuan, Trinh X., Magri, Christopher, \& Wadiak, James E. 1990, ApJS, 72, 245\\
Williams, Robert, \etal 1996 AJ, 112, 1335\\

\clearpage
\centerline{Figures}

Figure~\ref{fig:wfpc}.  HST WFPC2 image of UGC 12695 taken through the 814 (I band) filter 
(a), and the 300 (U band) filter (b) with a  2000s exposure time.
\vskip 0.2in
Figure~\ref{fig:gnd}.  Images of UGC 12695 taken using the MDM 1.3m telescope through the 
I, V, B, U and continuum subtracted H-\alp bands (Figure~\ref{fig:gnd} a - e, respectively). 
The images are 124'' across, with North up and East to the left.  Figure~\ref{fig:gnd}e also has the
regions of H-\alp emission studied in section 4 labeled.
See McGaugh (1992) for details on these images.
\vskip 0.2in
Figure~\ref{fig:sbprof}.  Surface brightness profile of UGC 12695 through the 814 filter.
The solid line is the surface brightness profile with all background galaxies
and H-\alp regions included in the analysis, the dashed line is UGC 12695 with
all the background galaxies (except U2-74, see text) removed, and the dash-dotted line is
UGC 12695 with both the background galaxies and all the H-\alp regions
removed. (See section 3.)
The errors in the surface brightness profiles are low, staying
below 0.20 \mss\ up to a radius of 40'', after which the errors
reach 0.30 - 0.50 \mss\ for the last 3 plotted points.  This error
comes primarily from uncertainties in the sky background.  Our
measured value is 21.6  counts/pixel with a mean error of 0.1 counts/per
pixel.  This maps onto a mean surface brightness through the 814 filter
of  21.5 \mss\ with a one sigma isophotal detection at 27.4
\mss\ (e.g. 0.1 counts/pixel).  Hence, we can do reliable isophotal detection
down to 6 magnitudes below sky with this data.

\vskip 0.2in
Figure~\ref{fig:nuc}.  Expanded view of the nucleus of UGC 12695 through the WFPC2 814 filter.
The image is 20'' x 20'', with the same orientation as Figure~\ref{fig:wfpc}(a).
\vskip 0.2in
Figure~\ref{fig:upict}.  Image showing the regions of H-\alp emission discussed in section 4.
The circled regions correspond to the H-\alp regions listed in
Table~\ref{tab:hcol}, while the white lines correspond to the spectral slits
used in McGaugh (1994) and presented in Table~\ref{tab:metal}.
\vskip 0.2in
Figure~\ref{fig:halp}.  Images of the H-\alp areas B and C discussed in section 4.
The image is 1'' across.  This image is the inverse (across the vertical axis) of
Figure~\ref{fig:upict}.
\vskip 0.2in
Figure~\ref{fig:backg}.  Images of the four largest background galaxies through the
814 filter.  The images are of U2-22, U2-23, U2-39, and U2-45 (Figure~\ref{fig:backg}a -- d,
respectively).

\clearpage
\centerline{Tables}

Table~\ref{tab:glob}.  The Global Properties of UGC 12695.
\vskip 0.2in
Table~\ref{tab:col}.  Colors of UGC 12695 Determined from the 814 and 300 WFPC2
Images.
\vskip 0.2in
Table~\ref{tab:hcol}.  Colors of the H-\alp Regions of UGC 12695.
\vskip 0.2in
Table~\ref{tab:metal}.  Metallicity of Select H-\alp Regions in UGC 12695 (McGaugh 1994).
\vskip 0.2in
Table~\ref{tab:halp}.  Photometry of Select Regions Within the H-\alp Areas B and C.
\vskip 0.2in
Table~\ref{tab:backg}.  Photometric and Structural Parameters Derived for the Background Galaxies.
\vskip 0.2in
Table~\ref{tab:gndgal}.  Colors of Eight of the Background Galaxies Identified in
the Ground Based (MDM) Images.

\clearpage
\setcounter{table}{0}
\begin{table}[t]
\begin{sideways}
\begin{tabular}{ccccccccccc}
\hline \hline
\multicolumn{11}{l}{}\\
&{\bf RA}& {\bf Dec}& {\bf V (km/s)}& {\bf d (Mpc)}& {\bf M$_{HI}$}&
{\bf M$_{dyn}$}& {\bf M$_{HI}$/L$_B$}& {\bf f${^B}_g$}& {\bf h (kpc)}& \\
&(1)&(2)&(3)&(4)&(5)&(6)&(7)&(8)&(9)\\
\multicolumn{11}{l}{}\\
&23:36:02.0& 12:52:32& 6182& 4590& 9.62& 9.98& 1.28& 0.62 & 8.4& \\
\multicolumn{11}{l}{}\\
\hline
\multicolumn{11}{l}{  }\\
&&{\bf M$_B$}& {\bf \Bmo}& {\bf m$_B$}& {\bf B-V}& {\bf U-B}& {\bf V-I}&\\
&&(10)&(11)&(12)&(13)&(14)&(15)\\
\multicolumn{11}{l}{}\\
&&-18.92& 23.8& 15.53& 0.26& -0.11& 0.95\\
\multicolumn{11}{l}{}\\
\hline
\hline
\end{tabular}
\end{sideways}
\caption{\label{tab:glob}}
\end{table}

\clearpage
\begin{table}[p]
\begin{sideways}
\begin{tabular}{cccccccccccccc}
\hline \hline
\multicolumn{11}{l}{}\\
& \multicolumn{3}{c}{\bf inner 2''}& & \multicolumn{3}{c}{\bf inner 10''}& & \multicolumn{3}{c}{\bf inner 30''}\\
\cline{2-4} \cline {6-8} \cline{10-12}
\multicolumn{11}{l}{}\\
&{\bf 814}& {\bf 300-814}& {\bf U-I}&&  {\bf 814}& {\bf 300-814}& {\bf U-I}&& {\bf 814}& {\bf 300-814}&
{\bf U-I}\\
a:& 22.215& -0.541& 2.159& & 19.426& -0.440& 2.260& & 18.069& -2.750& -0.092& \\
b:& 22.215& -0.541& 2.159& & 19.509& -0.977& 1.723& & 18.145& -2.879& -0.179& \\
c:& 22.215& -0.541& 2.159& & 19.511& -0.970& 1.730& & 18.145& -2.883& -0.183& \\
\multicolumn{11}{l}{}\\
\hline
\multicolumn{11}{l}{}\\
\multicolumn{11}{l}{Row a lists the colors with both the background galaxies and all H-\alp
regions included.}\\
\multicolumn{11}{l}{Row b lists the results of masking the background galaxies.}\\
\multicolumn{11}{l}{Row c lists the results of masking both the background galaxies and the H-\alp
regions.}\\
\multicolumn{11}{l}{All colors are within \plm 0.05.}\\
\multicolumn{11}{l}{}\\
\hline \hline
\end{tabular}
\end{sideways}
\caption{\label{tab:col}}
\end{table}
\clearpage
\begin{table}[p]
\begin{sideways}
\begin{tabular}{ccccccccc}
\multicolumn{9}{c}{}\\
\hline \hline
\multicolumn{9}{c}{}\\
{\bf Name}& {\bf RA }& {\bf Dec} & {\bf 300 - 814}& {\bf U-I}& R& 300$_T$& 814$_T$& R$_{IT}$\\
& (J2000) &(J2000)&&&('')&&&('')\\
\multicolumn{9}{c}{}\\
\hline
\multicolumn{9}{c}{}\\
A& 23:35:59.68& 12:52:31.4& -2.167& 0.533& 0.63& 21.194& 22.511& 1.09\\
B& 23:26:04.21& 12:52:25.5& -2.337& 0.363& 1.53& 19.284& 19.831& 2.15\\
C& 23:36:04.09& 12:52:29.4& -2.627& 0.077& 2.62& 17.964& 19.941& 3.31\\
D& 23:36:04.25& 12:52:39.9& -2.607& 0.093& 1.12& 20.374& 22.981& 1.12\\
E& 23:36:03.98& 12:52:46.5& -2.467& 0.223& 6.30& 17.514& 19.981& 6.30\\
F& 23:36:04.33& 12:52:47.3& -2.557& 0.143& 2.37& 19.724& 22.281& 2.37\\
\multicolumn{9}{c}{}\\
\hline
\multicolumn{9}{c}{}\\
\multicolumn{9}{l}{Column 6 lists the radius at which the colors were determined.}\\
\multicolumn{9}{c}{}\\
\hline \hline
\end{tabular}
\end{sideways}
\caption{\label{tab:hcol}}
\end{table}

\clearpage
\begin{table}[p]
\begin{center}
\begin{tabular}{cccc}
\multicolumn{4}{c}{}\\
\hline \hline
\multicolumn{4}{c}{}\\
{Region} & {log(O/H)} & {log (N/H)}& log(Ne/H) \\
\multicolumn{4}{c}{}\\
\hline
\multicolumn{4}{c}{}\\
{s1a1} & {-4.18} & {-6.52}& -4.84\\
{s1a2} & {-3.98} & {-5.66}& -4.76\\
{s1a3} & {-4.01} & {-5.81}& -4.86\\
{s2a1} & {-4.08} & {-5.81}& -4.83\\
{s2a2} & {-4.09} & {-5.62}& -4.96\\
s2a3   & -3.92   &  -     & -4.95\\
\multicolumn{4}{c}{}\\
\hline \hline
\end{tabular}
\end{center}
\caption{\label{tab:metal}}
\end{table}

\clearpage
\begin{table}[p]
\begin{center}
\begin{tabular}{cc}
\hline
\hline
\multicolumn{2}{c}{}\\
{Region 1:} &{300 - 814 = -2.499}\\
{Region 2:} &{300 - 814 = -2.481}\\
{Region 3:} &{300 - 814 = -2.950}\\
{Region 4:} &{300 - 814 = -2.555}\\
{Region 5:} &{300 - 814 = -2.633}\\
{Region 6:} &{300 - 814 = -5.076}\\
{Region 7:} &{300 - 814 = -2.537}\\
{Region 8:} &{300 - 814 = -1.820}\\
{inner 2'' (nucleus):} &{814 - 300 = -0.541}\\
{inner 10'' (nucleus):} &{814 - 300 = -0.977}\\
\multicolumn{2}{c}{}\\
\hline
\hline
\end{tabular}
\end{center}
\caption{\label{tab:halp}}
\end{table}

\clearpage
\vskip -3.0in
\begin{table}[p]
\begin{sideways}
\begin{tabular}{ccccccccccccc}
\multicolumn{13}{c}{}\\
\hline \hline
\multicolumn{13}{c}{}\\
{ Name}& { RA}& { Dec }& { 300-814}& { U-I}& 300$_T$& 814$_T$& 814$_T$(\alp)&
\alp ('') & \mo& \mo$_c$& R$_{T}$ ('')& {\it i}\\
(1)&(2)&(3)&(4)&(5)&(6)&(7)&(8)&(9)&(10)&(11)&(12)&(13)\\
\multicolumn{13}{c}{}\\
\hline
\multicolumn{13}{c}{}\\
U2-8&  23:36:01.64& 12:53:27.3& \gt -1.02& \gt 1.68& \gt 22.56& 23.581& 23.090& 0.380& 22.987& 23.487& 0.894& 50.9\\
U2-14& 23:35:58.22& 12:52:59.7& \gt -1.38& \gt 1.32& \gt 21.18& 22.561& 22.986& 0.246& 21.932& 22.263& 1.415& 42.5\\
U2-17& 23:36:02.66& 12:53:34.3& \gt -0.33& \gt 2.37& \gt 21.18& 21.511& 21.212& 0.632& 22.209& 22.719& 1.823& 51.3\\
U2-18& 23:35:58.48& 12:53:49.9& \gt 1.62& \gt 4.32&  \gt 21.18& 19.561& 18.164& 1.033& 20.223& 20.755& 4.127& 52.0\\
U2-20\dag& 23:36:02.06& 12:53:06.4& \gt -1.40& \gt 1.31& \gt 21.18& 22.571& 22.165& 0.539& 22.817& 22.878& 1.207& 19.1\\
U2-22\dag& 23:36:00.07& 12:52:36.2&  -0.867   & 1.833&      20.864& 20.991& 20.317& 0.800& 21.828& 21.891& 1.884& 19.3\\
U2-23\dag& 23:35:59.33& 12:52:40.1& -0.517    & 2.183&      21.444& 20.461& 18.694& 1.362& 21.360& 22.560& 4.277& 70.7\\
U2-24& 23:35:58.13& 12:53:56.5& \gt -2.89 & \gt -0.19&\gt 21.18& 24.071& 23.70& 0.297& 23.058& 23.911& 0.586& 62.9\\
U2-36\dag& 23:36:04.35& 12:52:59.9& \gt -0.95& \gt 1.75& \gt 21.19& 22.141& 21.220& 0.537& 21.864& 22.185& 1.270& 41.9\\
U2-39\dag& 23:36:03.58& 12:52:39.1& \gt 1.28& \gt 3.98&  \gt 21.19& 19.911& 18.475& 1.314& 21.063& 22.465& 5.596& 74.0\\
U2-41& 23:36:03.94& 12:53:04.1& \gt 0.92 & \gt 3.62& \gt 21.19& 20.271& 20.301& 0.086& 16.969& 17.564& 1.512& 54.5\\
U2-45\dag& 23:36:02.36& 12:52:41.0& \gt 0.70 & \gt 3.40& \gt 21.19& 20.491& 19.437& 0.863& 21.111& 21.971& 0.863& 63.1\\
U2-46\ddag& 23:36:06.68& 12:52:27.1& -& -& -& -& -& -& -& -& -& - \\
U2-48& 23:36:07.16& 12:53:04.7& \gt -1.23& \gt 1.47& \gt 21.19& 22.421& 22.634& 0.367& 22.515& 22.845& 0.367& 42.4\\
U2-53\ddag& 23:36:03.07& 12:53:37.1& -& -& -& -& -& -& -& -& -& - \\
U2-65& 23:36:01.84& 12:51:53.4& \gt -3.30& \gt -0.59&\gt 21.12& 24.411& 24.173& 0.199& 23.448& 22.667& 0.199& 60.9\\
U2-67& 23:36:04.74& 12:51:59.4& \gt -1.96& \gt 0.74& \gt 21.12& 23.081& 22.971& 0.408& 23.022& 23.830& 0.948& 61.6\\
U2-68& 23:36:04.27& 12:51:04.1& \gt -0.34& \gt 2.36& \gt 21.12& 21.461& 21.405& 0.363& 21.099& 21.654& 2.219& 53.1\\
U2-70& 23:36:04.21& 12:51:19.7& \gt -1.66& \gt 1.04& \gt 21.12& 22.781& 22.841& 0.206& 21.400& 21.593& 1.109& 33.2\\
U2-72& 23:36:01.22& 12:51:59.1& \gt -1.21& \gt 1.49& \gt 21.12& 22.331& 21.150& 0.726& 22.449& 22.927& 1.394& 29.9\\
U2-74& 23:36:02.09& 12:52:20.2& \gt 0.36& \gt 3.06&  \gt 21.12& 20.761& 21.283& 0.454& 21.565& 22.290& 2.050& 59.1\\
\multicolumn{13}{c}{}\\
\hline
\multicolumn{13}{c}{}\\
\multicolumn{13}{l}{\dag These galaxies were masked while determining the photometry of UGC 12695
(Table~\ref{tab:col}, rows b \& c).}\\
\multicolumn{13}{l}{Other than U2-74, the other galaxies in this table are located too far
from the nucleus of UGC 12695 to have been included}\\
\multicolumn{13}{l}{ in any of the WFPC2 photometry.
U2-74 was not masked as its identify as a background galaxy is unsure.}\\
\multicolumn{13}{c}{}\\
\multicolumn{13}{l}{\ddag Galaxy is on the edge of the WFPC2 image.  Photometry was not possible.}\\
\multicolumn{13}{c}{}\\
\hline\hline
\end{tabular}
\end{sideways}
\caption{\label{tab:backg}}
\end{table}

\clearpage
\vskip -3.0in
\begin{table}[p]
\begin{center}
\begin{tabular}{cccccc}
\multicolumn{6}{c}{ }\\
\hline \hline
\multicolumn{6}{c}{ }\\
&{\bf Name}& {\bf B-V}& {\bf V-I}& {\bf Radius ('')}&\\
\multicolumn{6}{c}{ }\\
\hline
\multicolumn{6}{c}{ }\\
&U2-17& -        & \gt 1.92& 1.82&\\
&U2-22& \gt 0.88 &     1.40& 1.10&\\
&U2-23& \gt 0.00 &     1.91& 0.95&\\
&U2-36& -        & \gt  1.29& 1.27&\\
&U2-39& \gta-0.12\dag &     0.16& 5.50&\\
&U2-41& \gt 0.49 &     3.07& 1.55&\\
&U2-45&     \dag &     1.75& 0.86&\\
&U2-72& \gt -1.34&     2.84& 1.40&\\
\multicolumn{6}{c}{ }\\
\hline
\multicolumn{6}{l}{ }\\
& \multicolumn{5}{l}{\dag Lies behind nucleus of UGC 12695.}\\
\multicolumn{6}{l}{ }\\
\hline \hline
\end{tabular}
\caption{\label{tab:gndgal}}
\end{center}
\end{table}

\clearpage
\begin{figure}[p]
\centerline{
\epsfxsize=7.0in
{See accompanying JPEG plate oneil.fig1a.jpeg}}
\caption{a\label{fig:wfpc}}
\end{figure}

\addtocounter{figure}{-1}
\clearpage
\begin{figure}[p]
\centerline{
\epsfxsize=7.0in
{See accompanying JPEG plate oneil.fig1b.jpeg}}
\caption{b}
\end{figure}

\begin{figure}[p]
\centerline{
\epsfxsize=7.0in
{See accompanying JPEG plate oneil.fig2a.jpeg}}
\caption{a\label{fig:gnd}}
\end{figure}

\addtocounter{figure}{-1}
\begin{figure}[p]
\centerline{
\epsfxsize=7.0in
{See accompanying JPEG plate oneil.fig2b.jpeg}}
\caption{b}
\end{figure}

\addtocounter{figure}{-1}
\begin{figure}[p]
\centerline{
\epsfxsize=7.0in
{See accompanying JPEG plate oneil.fig2c.jpeg}}
\caption{c}
\end{figure}

\addtocounter{figure}{-1}
\begin{figure}[p]
\centerline{
\epsfxsize=7.0in
{See accompanying JPEG plate oneil.fig2d.jpeg}}
\caption{d}
\end{figure}

\addtocounter{figure}{-1}
\begin{figure}[p]
\centerline{
\epsfxsize=7.0in
{See accompanying JPEG plate oneil.fig2e.jpeg}}
\caption{e}
\end{figure}

\clearpage
\begin{figure}[p]
\centerline{
\epsfxsize=6.0in
\epsffile{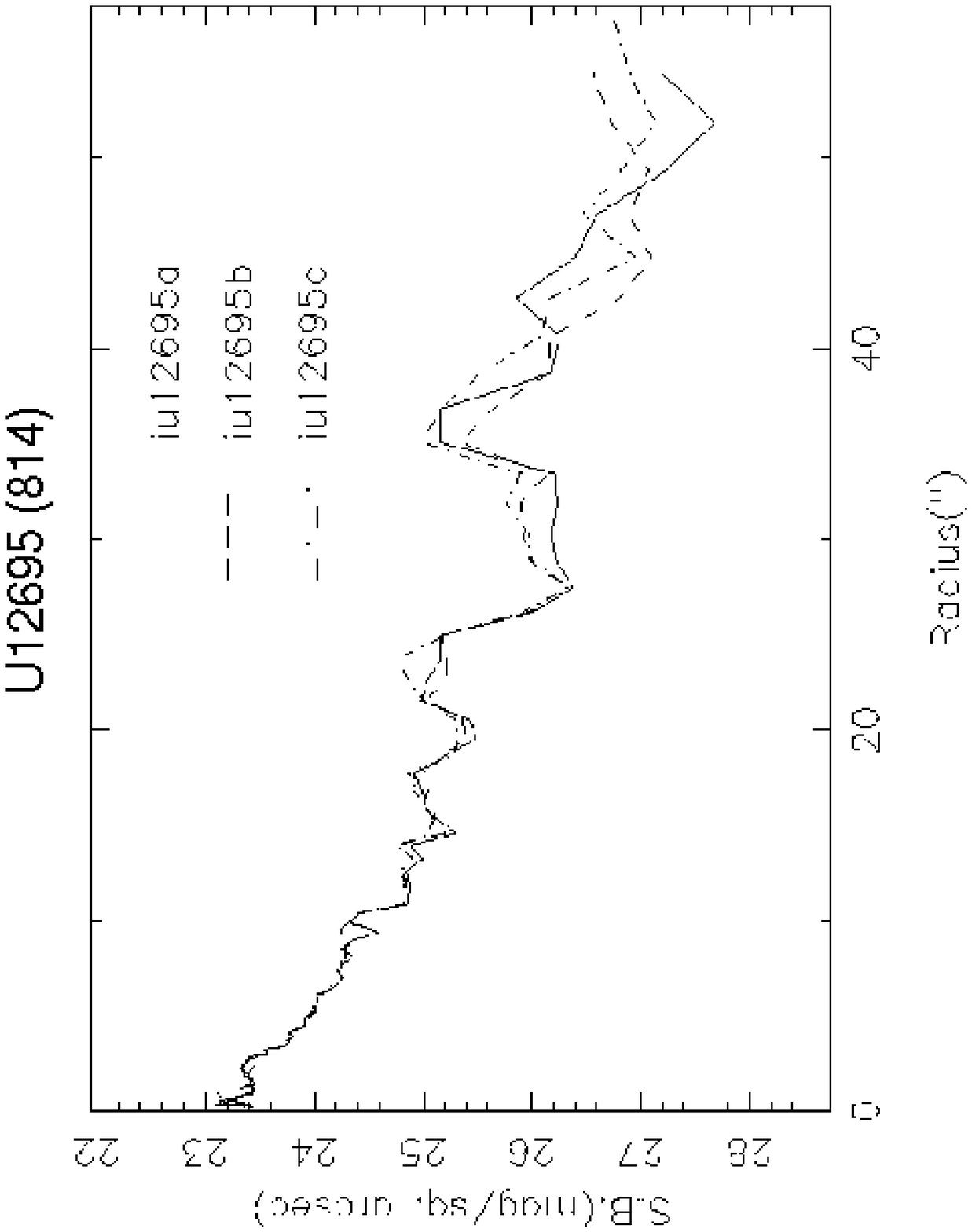}}
\caption{\label{fig:sbprof}}
\end{figure}

\clearpage
\begin{figure}[p]
\centerline{
\epsfxsize=7.0in
{See accompanying JPEG plate oneil.fig4.jpeg}}
\caption{\label{fig:nuc}}
\end{figure}

\clearpage
\begin{figure}[p]
\centerline{
\epsfxsize=7.0in
{See accompanying JPEG plate oneil.fig5.jpeg}}
\caption{\label{fig:upict}}
\end{figure}

\clearpage
\begin{figure}[p]
\centerline{
{See accompanying JPEG plate oneil.fig6.jpeg}}
\caption{\label{fig:halp}}
\end{figure}

\clearpage
\begin{figure}[p]
\centerline{
\epsfxsize=7.0in
{See accompanying JPEG plate oneil.fig7a.jpeg}}
\caption{\label{fig:backg}a}
\end{figure}

\clearpage
\addtocounter{figure}{-1}
\begin{figure}[p]
\epsfxsize=7.0in
\centerline{
{See accompanying JPEG plate oneil.fig7b.jpeg}}
\caption{b}
\end{figure}
 
\clearpage
\addtocounter{figure}{-1}
\begin{figure}[p]
\epsfxsize=7.0in
\centerline{
{See accompanying JPEG plate oneil.fig7c.jpeg}}
\caption{c}
\end{figure}
 
\clearpage
\addtocounter{figure}{-1}
\begin{figure}[p]
\epsfxsize=7.0in
\centerline{
{See accompanying JPEG plate oneil.fig7d.jpeg}}
\caption{d}
\end{figure}

\end{document}